# An Authentication Technique in Frequency Domain through Wavelet Transform (ATFDWT)

*Madhumita Sengupta, **J. K. Mandal, ***N. Ghoshal

*Computer Science & Engineering, Faculty of Engineering, Technology & Management

University of Kalyani

Kalyani, Nadia, West Bengal, India, (madhumita.sngpt@gmail.com)

**Computer Science & Engineering, Faculty of Engineering, Technology & Management

University of Kalyani

Kalyani, Nadia, West Bengal, India, (jkm.cse@gmail.com)

***Department of Engineering &Technological Studies, Faculty of Engineering, Technology & Management, University of Kalyani

Kalyani, Nadia, West Bengal, India, (nabin_ghoshal@yahoo.co.in)

## Abstract

In this paper a DWT based steganography in frequency domain, termed as ATFDWT has been proposed. Here, the cover image is transformed into the time domain signal through DWT, resulting four sub-image components as 'Low resolution', 'Horizontal orientation', 'Vertical orientation' and 'Diagonal orientation'. The secret message/image bits stream in varying positions are embedded in 'Vertical orientation sub-image' followed by reverse transformation to generate embedded/encrypted image. The decoding is done through the reverse procedure.

The experimental results against statistical and visual attack has been computed and compared with the existing technique like IAFDDFTT[1], in terms of Mean Square Error (MSE), Peak Signal to Noise Ratio (PSNR), Standard Deviation(SD) and Image Fidelity(IF) analysis, which shows better performances in ATFDWT.

# 1. Introduction

Communication over Internet has become incredibly popular, digital media are extensively used for transferring information. Steganography is an important technique to protect delicate information during transmission through public networks.

Mitchell D, et al., in the year 1996 proposed a technique of robust data hiding for images with encoding extra information in an image by making small modification to its pixels [5], the first method describes spatial masking and data spreading to hide information by modifying image coefficients. The second method uses frequency masking to modify image spectral components. Mauro Barni proposed a solution to the problem of copyright protection of multimedia data in a networked environment by digital watermarking [6], it makes possible to firmly associate a digital document with a code allowing the identification of the data creator, owner, authorized consumer, and so on. In the year 2005 Rongrong Ni, et al., proposed a semi-blind technique [7] to authenticate through embedding the image with a binary copyright symbol followed by Arnold iteration transform for constructing the watermark to increase the security. The secure image adaptive watermark is embedded in the feature blocks by modifying DCT middle-frequency coefficients. This detection and extraction is a semi-blind, because it does not need the original image but those who have the original watermark and the key, can detect and extract the right watermark, which provide high security level. In 2007 Guillaume Lavoué presented a non-blind watermarking scheme for subdivision surfaces [8]. Chin-Chen Chang in the same year proposed reversible hiding in DCT-based compressed images [9], in this scheme, the two successive zero coefficients of the medium-frequency components in each block were used to hide the secret data, and the scheme modifies the quantization table to maintain the quality of the stego-image. In 2008 Haohao Song, et al., proposed a contourlet based adaptive technique that decomposes an image into low-frequency (LF) sub-band and a high-frequency (HF) sub-band by Laplacian pyramid (LP). In this scheme, the LF sub-band is created by filtering the original image with 2-D low-pass filter and the HF sub-band is created by subtracting the synthesized LF sub-band from the original image, then secret message/image embedded into the contourlet coefficients of the largest detail sub-bands of the image.

IAFDDFTT [1] has been proposed using discrete Fourier transformation for image authentication, and in the same year Santa Agreste, et al., suggested an approach of pre-processing digital image for wavelet-based watermark [10] of colour image protection which is robust but not blind.

This paper proposes a frequency domain based technique termed as ATFDWT where the source image is transformed into its corresponding frequency domain and the authenticated messages/images are embedded into the frequency components of DWT. A reverse transform is performed as a final step of embeddings which added an extra layer of security to the process.

Various parametric tests are performed and results obtained are compared with existing IAFDDFTT, based on Mean Square Error (MSE), Peak Signal to Noise Ratio (PSNR), Standard Deviation (SD), and Image Fidelity (IF) analysis [4] to show a consistent relationship with the quality perceived by the HVS.

Section 2 deals with the proposed technique. Section 3 analyzed the results of proposed work and comparisons to with existing technique are presented in section 4. Conclusions are drawn in section 5 and references are cited at the end.

## 2. The Technique

ATFDWT is divided into four phases. Forward transformation of cover image is performed at the beginning to convert the image from spatial domain to frequency using DWT, which is computed by successive lowpass and highpass filtering of the discrete time-domain signal. After transformation, embedding phase runs to embed 't' number of bits into each bytes of cover image in varying positions selected through hash function. Fidelity adjustment is done on embedded image to improve the quality of the image without hampering the inserted information bits. A reverse transformation is performed as final step of embedding which generate the embedded image. At destination, the received image passes through forward transformation using DWT and based on same hash function, secret message/image bits are extracted to reconstruct the message/image. On comparing extracted secret message/image with original secret message/image, the received image at destination is authenticated. The schematic diagram of ATFDWT is given in figure 1, and details of each stage described in section 2.1 to 2.4.

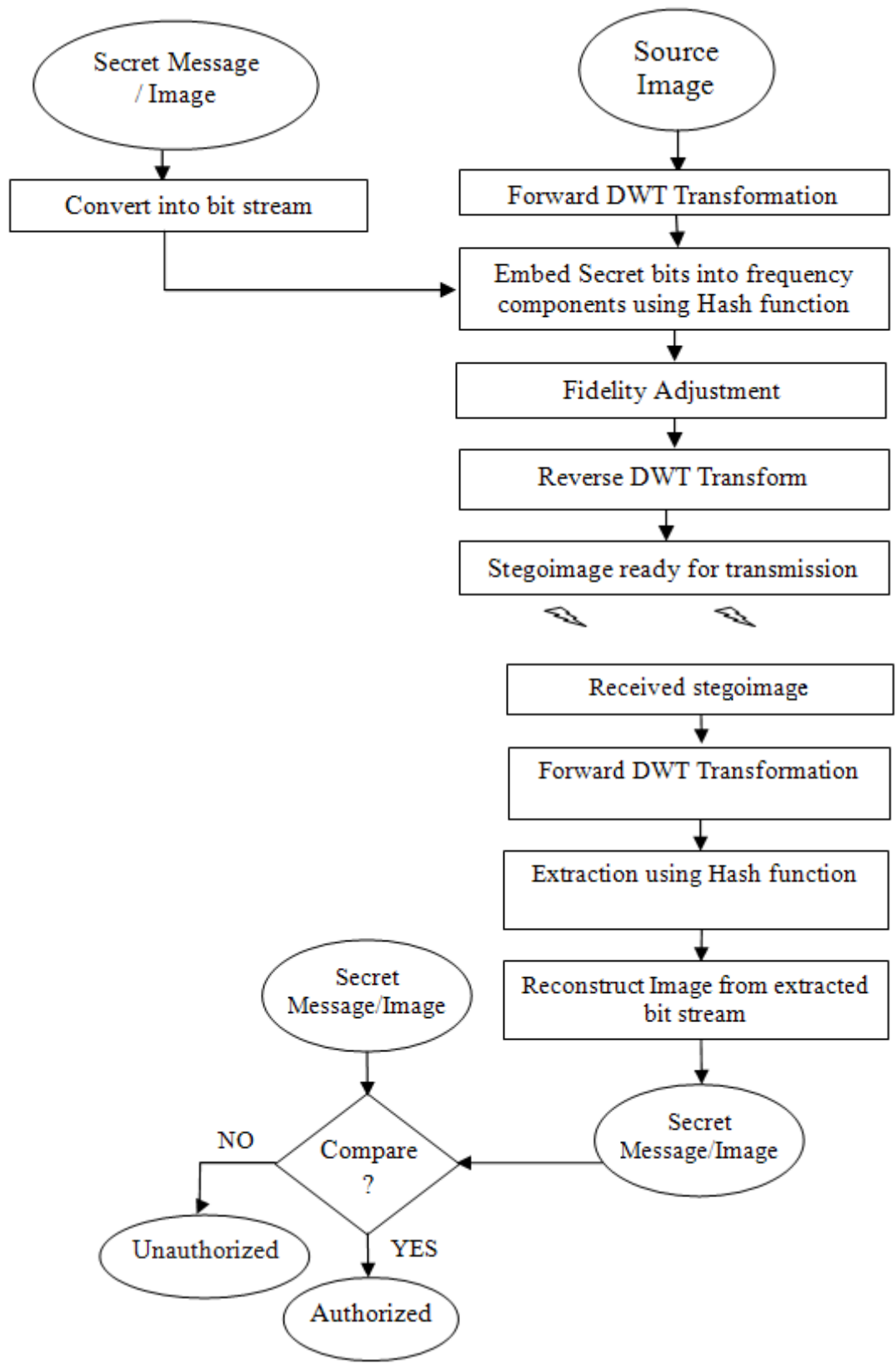

Fig. 1: Schematic diagram of ATFDWT

## 2.1 Transformation

Wavelet based forward transformation [11] converts image from spatial domain to frequency domain using eq (1) and eq (2), and that of inverse transformation is guided by eq.(3). Mathematically, the image matrixes multiply with scaling function coefficients and wavelet function coefficients in the formation of transformation matrix [12] (fig 3.b, fig 3.e).

$$Y_{Low}[k] = \sum_n x[n].h[2k-n] \qquad (1)$$

$$Y_{High}[k] = \sum_n x[n].g[2k-n] \qquad (2)$$

$$x[n] = \sum_{k=-\infty}^{\infty}\bigl(Y_{High}[k].g[2k-n]\bigr) + (Y_{Low}[k].h[2k-n]) \qquad (3)$$

Where x[n] is original signal, h[x] is half band low pass filter, g[x] is half band high pass filter, $Y_{Low}[k]$ is output of high pass filter after sub sampling by 2, $Y_{High}[k]$ is output of low pass filter after sub sampling by 2.

Forward transformation is discussed in section 2.1.1 and that of inverse transformation is in section 2.1.2.

### 2.1.1 Forward Transformation

In the proposed technique *Mallat* based two-dimensional wavelet transform is used in order to obtain a set of bi-orthogonal subclasses of images [3]. In two-dimensional wavelet transformation, a scaling function φ(x, y) represent by eq (4).

$$\varphi\,(x,\,y) = \varphi(x)\,\varphi(y) \qquad (4)$$

If ψ(x) is a one-dimensional wavelet function associated with one–dimensional scaling function φ(x), three two dimensional wavelets may be defined as given in eq (5). Fig 2 represents functions in visual form.

$$\left.\begin{array}{l} \psi^H(x,y) = \varphi\,(x)\,\psi(y) \\ \psi^V(x,y) = \psi\,(x)\,\varphi(y) \\ \psi^D(x,y) = \psi\,(x)\,\psi(y) \end{array}\right\} \qquad (5)$$

|  |  |
|---|---|
| *Low resolution sub-image* $\psi(x,y)= \varphi(x)\varphi(y)$ | *Horizontal orientation sub-image* $\psi^H(x,y)= \varphi(x)\psi(y)$ |
| *Vertical orientation sub-image* $\psi^V(x,y)= \psi(x)\varphi(y)$ | *Diagonal orientation sub-image* $\psi^D(x,y)= \psi(x)\psi(y)$ |

Fig. 2: Image decomposition through DWT

Consider a 4 x 4 matrix fig 3.a with pixel intensity values taken from Monalisa image (fig 7.a) in PPM format and that of row transformation matrix in fig 3.b. to generate row transformed matrix(fig 3.c) through convolution operation. In the second step of operation, generated row transformed matrix(fig 3.c) is taken as input and convoluted with column transformation matrix(fig 3.d) to generate column transformed matrix(fig 3.e) which is the output of one round of forward transformation technique and is given in fig.3.f. on the basis of four frequency quarter. The detail computation is discussed in section 2.1.3.

| 191 | 187 | 206 | 198 |
|---|---|---|---|
| 171 | 151 | 186 | 186 |
| 130 | 106 | 116 | 168 |
| 112 | 120 | 136 | 140 |

Fig. 3.a. Image matrix

| $H_0$ | 0 | $G_0$ | 0 |
|---|---|---|---|
| $H_1$ | 0 | $G_1$ | 0 |
| 0 | $H_0$ | 0 | $G_0$ |
| 0 | $H_1$ | 0 | $G_1$ |

3.b. Row transformation matrix

| 189 | 202 | 2 | 4 |
|---|---|---|---|
| 161 | 186 | 10 | 0 |
| 118 | 142 | 12 | -26 |
| 116 | 138 | -4 | -2 |

3.c. Row transformed matrix

| $H_0$ | $H_1$ | 0 | 0 |
|---|---|---|---|
| 0 | 0 | $H_0$ | $H_1$ |
| $G_0$ | $G_1$ | 0 | 0 |
| 0 | 0 | $G_0$ | $G_1$ |

3.d. Column transformation matrix

| 175 | 194 | 6 | 2 |
|---|---|---|---|
| 117 | 140 | 4 | -14 |
| 14 | 8 | -4 | 2 |
| 1 | 2 | 8 | -12 |

3.e. Column transformed matrix

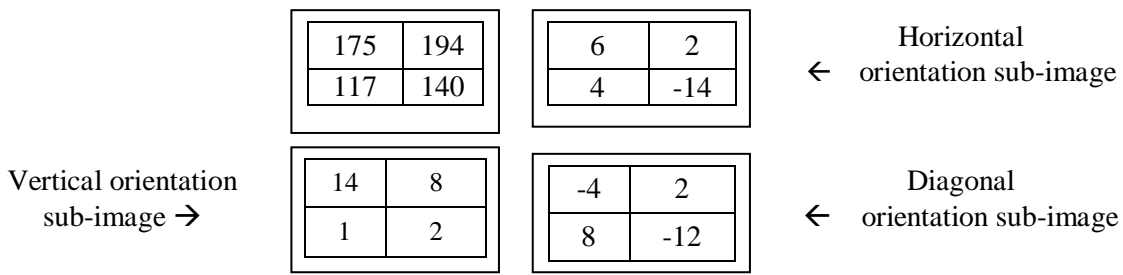

3.f. Outputs of forward transformation technique

Fig. 3: Example of image decomposition in Mallet Wavelet

As per Haar forward transform, scaling function coefficients and wavelet function coefficients [12] are taken as $H_0 = ½$, $H_1 = ½$, $G_0 = ½$ $G_1 = -½$.

### 2.1.2 Inverse Transformation

Inverse transformation is just the reverse of the forward transformation and is also reversible. Here column transformation is performed first followed by row transformation. But the coefficient values are different for column/row transformation matrices. The coefficient for reverse transformations are $H_0 = 1$, $H_1 = 1$, $G_0 = 1$, $G_1 = -1$ [12].

### 2.1.3 Transformation Computation

Transformations are performed through simple addition and division. Consider the matrix(fig 3a) given in figure 3 as source image matrix, values of first 2 x 2 sub matrix are 191, 187, 171, 151 [representing $X_{ij}$, where $i$ and $j$ are row and column respectively]. If forward transformation is applied, the resultant frequency components will be 175, 6, 14, -4, incorporating low resolution, horizontal orientation, vertical orientation and diagonal orientation sub images respectively. The computation for forward transformation sub matrices is as follows.

For low resolution sub image $X'_{ij}$ will be $\sum X_{ij} / 4 \Rightarrow (X_{00} + X_{01}) + (X_{10} + X_{11}) / 4$. Thus as per example it is $[(191 + 187) + (171 + 151)] / 4 \Rightarrow [378 + 322] / 4 \Rightarrow 175$ (Lr low resolution). For horizontal orientation sub image $X'_{ij}$ will be $[(X_{00} - X_{01}) + (X_{10} - X_{11}) / 4 \Rightarrow [(191 - 187) + (171 - 151)] / 4 \Rightarrow [4 + 20] / 4 \Rightarrow 6$ (H$_o$ horizontal orientation). For vertical orientation sub image it is $(X_{00} + X_{01}) - (X_{10} + X_{11}) / 4 \Rightarrow [(191 + 187) - (171 + 151)] / 4 \Rightarrow [378 - 322] / 4 \Rightarrow 14$ (V$_o$ vertical orientation). For diagonal orientation sub image it is $(X_{00} - X_{01}) - (X_{10} - X_{11}) / 4 \Rightarrow [(191-187)-(171-151)] / 4 \Rightarrow ([4 - 20])/4 \Rightarrow -4$ (D$_o$ diagonal orientation).

In case of inverse transformation, as the coefficient value ranges from 1 to -1 no division is required, only addition and subtraction is enough to compute back from frequency domain to spatial domain. The computations to generate the 2 x 2 original image matrixes $X_{ij}$ for the coordinate $X_{00} = (L_r + H_o) + (V_o + D_o) = (175 + 6) + (14 + (-4)) \Rightarrow 191$, $X_{01} = (L_r - H_o) + (V_o - D_o) \Rightarrow (175 - 6) + (14 - (-4)) \Rightarrow 187$, $X_{10} = (L_r + H_o) - (V_o + D_o) \Rightarrow (175 + 6) - (14 + (-4)) \Rightarrow 171$, $X_{11} = (L_r - H_o) - (V_o - D_o) \Rightarrow (175 - 6) - (14 - (-4)) \Rightarrow 151$. Thus in general form for a mask of $2 \times 2$ on image, forward transformation become, $\sum ((X_{ij} + X_{i,j+1}) + (X_{i+1,j} + X_{i+1,j+1})) / 4$, $\sum ((X_{ij} - X_{i,j+1}) + (X_{i+1,j} - X_{i+1,j+1})) / 4$, $\sum ((X_{ij} + X_{i,j+1}) - (X_{i+1,j} + X_{i+1,j+1})) / 4$, and $\sum ((X_{ij} - X_{i,j+1}) - (X_{i+1,j} - X_{i+1,j+1})) / 4$, for low resolution, horizontal, vertical and diagonal orientation. For inverse transformation, the general form become, $X_{ij} = (L_r + H_o) + (V_o + D_o)$, $X_{i,j+1} = (L_r - H_o) + (V_o - D_o)$, $X_{i+1,j} = (L_r + H_o) - (V_o + D_o)$ and $X_{i+1,j+1} = (L_r - H_o) - (V_o - D_o)$.

The complexity of transformation of the matrix containing the dimension M x N is O(MN).

## 2.2 Embedding Technique

On first level forward transformation through DWT the image is represented in four sub images[3] as shown in fig.2. In proposed ATFDWT 'vertical orientation sub-image' is taken to hide two bits per bytes of vertical orientation sub-image. The position is selected on the basis of hash function, which is discussed in section 2.2.1.

### 2.2.1 Insertion

Bits are inserted based on a hash function where embedding position in cover image are selected using formula (K%S) and (K%S) +1 where K varies from 0 to 7 and S from 2 to 7. For example, if the vertical orientation sub-image representing values as shown in fig. 4, with 8bit representation, 32bits of information as bits stream S$ = "10011001, 11100101, 10011101, 11001101". Data is hidden in varying positions selected by hash function up to LSB+3 as given in fig. 5, where bold bits are embedded information.

| 65 | 78 | 73 | 30 |
|----|----|----|----|
| 58 | 78 | 38 | 32 |
| 56 | 73 | 56 | 35 |
| 59 | 70 | 52 | 39 |

→

| 01000001 | 01001110 | 01001001 | 00011110 |
|----------|----------|----------|----------|
| 00111010 | 01001110 | 00100110 | 00100000 |
| 00111000 | 01001001 | 00111000 | 00100011 |
| 00111011 | 01000110 | 00110100 | 00100111 |

Fig. 4: Pixel intensity/binary values of vertical orientation sub-image.

| 010000**01** | 01001**100** | 0100**0**101 | 0001**0**111 |
|----------|----------|----------|----------|
| 001110**11** | 01001**010** | 0010**1**010 | 0010**0**001 |
| 001110**01** | 01001**101** | 0011**1**100 | 0010**0**011 |
| 001110**11** | 01000**000** | 0011**1**100 | 0010**0**111 |

→

| 65 | 76 | 69 | 23 |
|----|----|----|----|
| 59 | 74 | 42 | 33 |
| 57 | 77 | 60 | 35 |
| 59 | 64 | 60 | 39 |

Fig. 5: Embedded image matrix after embedding S$

| 010000**01** | 01001**101** | 0100**0**111 | 0010**0**001 |
|----------|----------|----------|----------|
| 001110**11** | 01001**011** | 0010**1**000 | 0010**0**001 |
| 001110**01** | 01001**100** | 0011**1**100 | 0010**0**011 |
| 001110**11** | 01001**000** | 0011**1**100 | 0010**0**111 |

→

| 65 | 77 | 71 | 33 |
|----|----|----|----|
| 59 | 75 | 40 | 33 |
| 57 | 76 | 60 | 35 |
| 59 | 72 | 60 | 39 |

Fig. 6: Embedded image embedded with S$ along with fidelity adjustment

The Extraction technique is just the reverse of insertion.

## 2.3 Fidelity adjustment

Adjustment of fidelity is an added step to enforce minimum changes in image after embedding hidden data without hampering the bits inserted, that minimizes the deviation of fidelity, in bit level. For two bits of hidden data per byte of cover image, four cases may exist, case I when position for two bits is (LSB and LSB+1), case II for position (LSB+1 and LSB+2), case III position (LSB+2 and LSB+3) and case IV for position (LSB+3 and LSB). In each case, 12 possibilities will occur. So far as permutation is concerned, it allows 16 cases, but the cases are taken into account where difference of embedded and original pixel value is not equal to zero.

Let cover image pixel value is $C_i$, and $C_i$` is the pixel value after embedding (stegoimage pixel value), secret message/image pixel value $S_i$, difference value $D_i$, Position of embedding $P_1$ and $P_2$ for two bits per byte of cover image and byte representation is ($b_7$, $b_6$, $b_5$, $b_4$, $b_3$, $b_2$, $b_1$, $b_0$) 8 bits representation, where $b_0$ is LSB and $b_7$ is MSB.

Case I: The position $P_1$ and $P_2$ are LSB and LSB+1 that is for $C_i$, $b_0$ and $b_1$, on embedding $S_i$'s $b_7$ and $b_6$ $C_i$ become $C_i$` (stegoimage pixel value), the range of differences ( $D_i = C_i - C_i$` ) can pass through from +1, -1, +2, -2, +3 and -3, only +/-3 needed to be handled.

If $C_i$ is greater than 3 and difference is +/-3 then $C_i$` needed to handle FA technique else no change is required. For example, $C_i = 15$, $S_i$'s $b_7 = $ '0' & $b_6 = $ '0', then $C_i$`= 12 thus, the difference become $D_i = $ (-3). Then after handling, $C_i$` become 16 (10000), and the new difference become +1.

Case II : The position $P_1$ and $P_2$ are LSB+1 and LSB+2 that is for $C_i$ , $b_1$ and $b_2$ on embedding $S_i$'s $b_5$ and $b_4$ $C_i$ become $C_i$` (stegoimage pixel value), the range of differences ( $D_i = C_i - C_i$` ) will be restricted to +2, -2, +4, -4, +6 and -6.

If $C_i$ is greater than 6 and the difference is +/- 6 or +/- 4 then $C_i$` needed to be handled through FA technique, else directly convert $b_0$ to '0' for $D_i$ equal to -2 and $b_0$ to '1' for $D_i$ equal to +2.

Case III: The position $P_1$ and $P_2$ are LSB+2 and LSB+3 that is for $C_i$ , $b_2$ and $b_3$ on embedding $S_i$'s $b_3$ and $b_2$ $C_i$ become $C_i$` (stego-image pixel value), the range of differences ( $D_i = C_i - C_i$` ) will be restricted to +4, -4, +8, -8 +12, and -12.

If $C_i$ is greater than 15 then $C_i$` needed to be handled through FA technique.

Case IV: The position $P_1$ and $P_2$ are LSB+3 and LSB+4 that is for $C_i$ $b_3$ and $b_4$ on embedding $S_i$'s $b_1$ and $b_0$ $C_i$ become $C_i$` (stego-image pixel value), the range of differences ( $D_i = C_i - C_i$` ) will be restricted to +1, -1, +7, -7, +8, -8, +9 and -9. If $D_i$ is +/- 7 or +/- 8 or +/-9 then

$C_i$` needed to be handled through FA technique, else no change is required. The algorithm is given in section 2.3.1.

## 2.3.1 Algorithm

For fidelity adjustment separate cases are defined. If the difference in value is negative, then we need to adjust in such a manner that the existing value comes down, if the difference in value come positive, then it needed to be handled to increase the value.

Consider two embedded positions are $P_1$ and $P_2$.

Step 1: Calculate the difference $d_i$ ($d_i = C_i - C_i$`).

Step 2: If $d_i < 0$ then, for $C_i$` suppose two embedded positions are $b_i$ and $b_{i+1}$ then check the bit value of $b_{i+2}$ if it is '0' then check $b_{i+3}$ and so on up to $b_7$, on receiving of first bit value '1' flip it to '0' and then flip all the 0's to '1' up to $b_{i+2}$ and from $b_{i-1}$ to $b_0$. Not to change $b_i$ and $b_{i+1}$ as these two are embedded with secret message/image.

Step 3: If $d_i > 0$ then, for $C_i$` suppose two embedded positions are $b_i$ and $b_{i+1}$ then check the bit value of $b_{i+2}$ if it is '1' then check $b_{i+3}$ and so on up to $b_7$, on receiving of first bit value '0' flip it to '1' and then flip all the 1's to '0' up to $b_{i+2}$ and from $b_{i-1}$ to $b_0$. Not to change $b_i$ and $b_{i+1}$ as these two are embedded with secret message/image.

Step 4: If $d_i = 0$ then, no change is required, original bit value will roll back.

In general, on embedding two secret bits, in position $b_i$ and $b_{i+1}$ within lower nibble, where i varies from 0 to 3, handling is required to optimize the intensity by adjusting upper bits from $b_{i+2}$ toward MSB. If difference $d_i$ become less than 0 then the first $k^{th}$ bit value 1 from $b_{i+2}$ towards MSB needed to be flipped to zero, and rest all zeros from $b^{k-1}$ to $b^0$ needed to be flipped to 1 except $b_i$ and $b_{i+1}$. Else, if difference $d_i$ become greater than 0 then the first $k^{th}$ bit value 0 from $b_{i+2}$ towards MSB needed to be flipped to 1, and the rest all ones from $b^{k-1}$ to $b^0$ need to be flipped to zero except $b_i$ and $b_{i+1}$.

Here amount of intensity of pixel escalated due to flipping of bit value 1 to zero, is $2^k$, and all bits from $2^{k-1}$ towards $2^0$ excluding $b_i$ and $b_{i+1}$ flipped from zero to 1 thus, intensity value become $2^k$ verses $\sum_{n=1}^{k} 2^{k-n} - 2^i - 2^{i+1}$ where k is the flipped bit position and i is the first embedding position. In other words, if for adjustment, we increases the intensity by $2^k$ then simultaneously we are decreasing the value by $\sum_{n=1}^{k} 2^{k-n} - 2^i - 2^{i+1}$ that must be less than $2^k$, and the amount of value originally changed on embedding $b_i$ and $b_{i+1}$ is optimized.

Adjustment technique works on bit level, based on various ranges of difference in value ($d_i$), in total twelve possibilities in each four cases (4 x 12) are possible. Let's consider the original value of pixel before embedding is 224 (11100000), suppose LSB+2 and LSB+3 are to be replaced by secret bits. In worst case the new value after embedding is 236 (11101100) and difference is -12, then on handling FA the value come down to 223 where the difference after FA will be 1 without changing the embedded bits at position LSB+2 and LSB+3. Fig 6 shows the fidelity adjustment on matrix shown in fig 5.

## 2.4 Complexity

Complexity of the proposed ATFDWT will be $O(MN)$ where M represents number of rows, and N represents number of column. The step by step procedure is shown below.

Here, the cover image of dimension M x N, and secret image of dimension is used as input, and the stego image of dimension M x N will be the output.

Step 1: To read the color image of (M x N) the complexity $\left(\frac{M}{4} \times \frac{N}{4}\right)$ will be $O(M \times N)$.

Step 2: Complexity to transform image from spatial domain to frequency domain as stated in section (2.1.3 Transformation Computation) is $O(M \times N)$.

Step 3: DWT may obtain floating point values, to convert floating values to integer values complexity will be $O(M \times N)$.

Step 4: If we fetch any single frequency components out of four, to do so, the complexity will be, $O\left(\frac{M}{2} \times \frac{N}{2}\right)$ in case we fetch two frequency components, the complexity will be $O\left(\frac{M}{2} \times \frac{N}{2}\right) + O\left(\frac{M}{2} \times \frac{N}{2}\right)$, in any of the cases the complexity cannot be more than $O(M \times N)$.

Step 5: The secret image taken as compared to original cover image is ¼ $^{th}$ thus, to read the secret image, the complexity will be $O\left(\frac{M}{4} \times \frac{N}{4}\right)$.

Step 6: To convert all the pixels of secret image into binary bits the complexity will be $O\left(\frac{M}{4} \times \frac{N}{4}\right) \times 8$, which imply $O\left(\frac{M \times N}{2}\right)$.

Step 7: From 1/4$^{th}$ cover image frequency components, per pixel two bits are selected to swap with secret bits of step 6, with respect of position vector K, where K = 0 to 6, the complexity will be in the range of $O(M \times N) \times K$.

Step 8: Inverse transformation is performed with complexity $O(M \times N)$.

The resultant complexity of the proposed algorithm is $O(M \times N) + O(M \times N) + O(M \times N) +$ , $O(M \times N) + O\left(\frac{M}{4} \times \frac{N}{4}\right) + O\left(\frac{M \times N}{2}\right) + O(M \times N) \times K + O(M \times N)$ where k is constant.

Thus the complexity of the ATFDWT will be $O(M \times N)$.

# 3    Results and Discussions

This section deals with the result of computation after embedding with hidden data. Ten PPM [2] images have been taken and ATFDWT is applied on each. All cover images are 512 x 512 in dimension and a gold coin (k) of 128x128 is used as authenticating image. The images along with the results of standard deviation are given in fig. 7, average of MSE for 10 images are 3.694128 and PSNR is 42. 479033 and image fidelity is 0.999652, and standard deviation is 52.866214 as given in table 1.

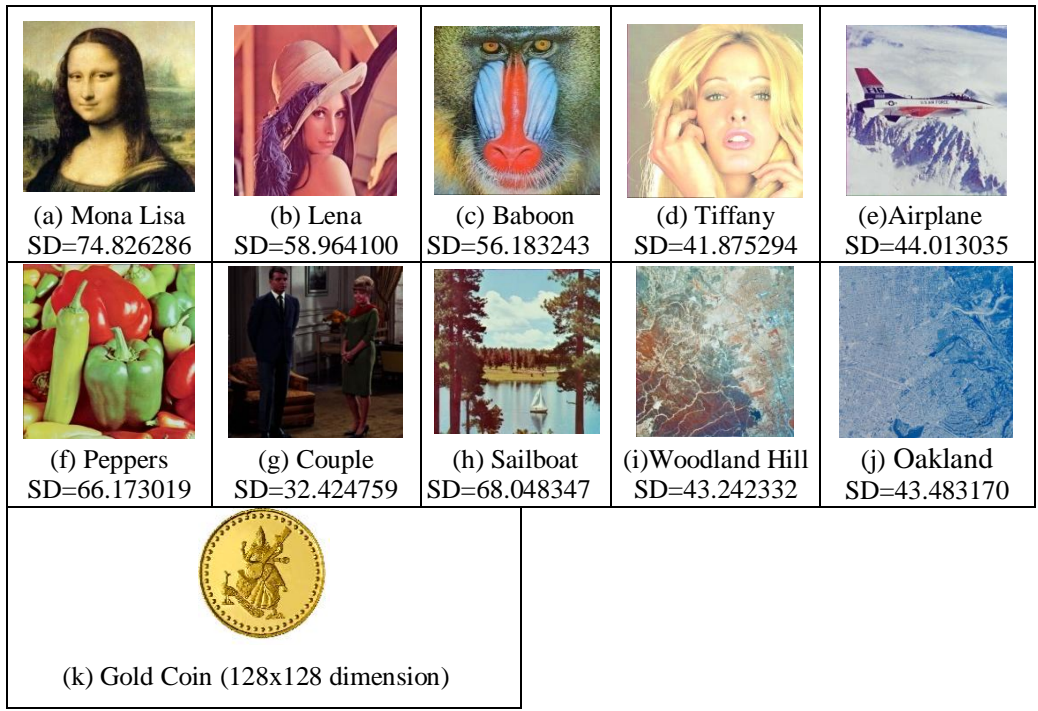

Fig. 7: Cover images of dimension 512x512 and that of secret image 128x128

The results of MSE, PSNR, SD and IF for each of ten benchmark images on embedding gold coin (49152 bytes) 2bits per byte of cover image in the position selected by hash function is given separately. All benchmark images obtain very high PSNR and IF values except in case of Airplane and Couple, the values are marginally low. The results obtained for all ten benchmark images in terms of all four attributes are better than existing technique.

Table 1:  MSE, SD, PSNR and IF in embedding 2 bits of hidden data.

| | Cover Image 512 x 512 | MSE | PSNR | SD for Stegoimage | IF |
|---|---|---|---|---|---|
| (a) | Mona Lisa | 3.293662 | 42.954014 | 74.755318 | 0.999816 |
| (b) | Lena | 3.885129 | 42.236749 | 58.960365 | 0.999805 |
| (c) | Baboon | 3.188728 | 43.094628 | 56.093765 | 0.999834 |
| (d) | Tiffany | 3.999738 | 42.110488 | 41.921524 | 0.999906 |
| (e) | Airplane | 4.109852 | 41.992542 | 43.992962 | 0.999883 |
| (f) | Peppers | 3.886759 | 42.234927 | 66.036728 | 0.999766 |
| (g) | Couple | 4.343258 | 41.752647 | 32.373444 | 0.997985 |
| (h) | Sailboat | 3.580195 | 42.591737 | 68.040909 | 0.999820 |
| (i) | Woodland Hill | 3.255777 | 43.004257 | 43.152962 | 0.999864 |
| (j) | Oakland | 3.398177 | 42.818344 | 43.334167 | 0.999841 |
| | *Average* | *3.694128* | *42.479033* | *52.866214* | *0.999652* |

## 4  Comparison with IAFDDFTT

In this section, a comparative study has been made between IAFDDFTT [1] and ATFDWT in terms of mean square error, peak signal to noise ratio, standard deviation, and image fidelity. Comparison is done on the ten PPM images given in fig. 7 of section 3. Gold coin is embedded as secret image of dimension 128x128. Two bits, from secret bits stream are embedded in two random positions up to LSB+3 in cover images based on hash function. Outputs are shown in fig.8.

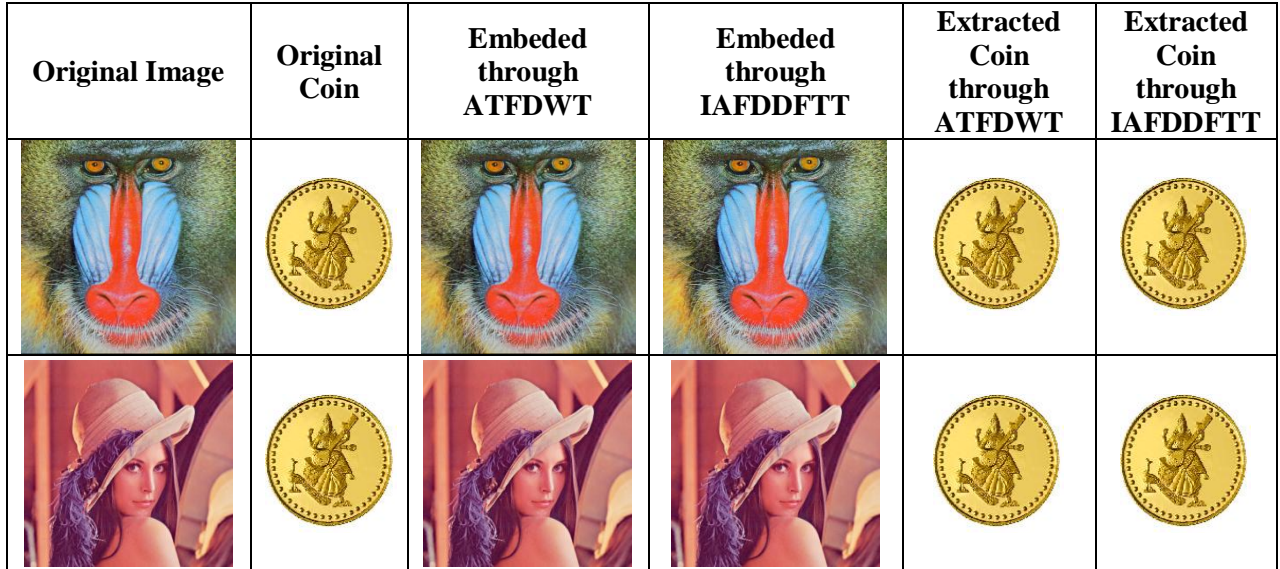

Fig. 8: Comparison of images after embedding Gold coin in different images using ATFDWT and IAFDDFTT

### 4.1  Error Rate (MSE)

In statistics, the mean square error or MSE [4] of an estimator is one of the many ways to quantify the difference between an estimator and the true value of the quantity being estimated. The MSE represents the cumulative squared error between the embedded and the original image, the lower the value of MSE, the lower the error. On comparison with the existing technique ATFDWT reveal better performances. The minimum difference of MSE is 0.599429(Peppers), and of maximum difference is 1.392843 [Tiffany (5.392581-3.999738)] as shown in table 2. Graphical representation of the same is given in fig. 9, which exhibits better performances for the proposed ATFDWT.

Table 2: Comparison of MSE between IAFDDFTT and ATFDWT

| IMAGES 512 x 512 | MSE | |
| --- | --- | --- |
| | IAFDDFTT | ATFDWT |
| Mona Lisa | 4.498556 | 3.293662 |
| Lena | 4.490079 | 3.885129 |
| Baboon | 4.116379 | 3.188728 |
| Tiffany | 5.392581 | 3.999738 |
| Airplane | 4.907283 | 4.109852 |
| Peppers | 4.486188 | 3.886759 |
| Couple | 5.207868 | 4.343258 |
| Sailboat | 4.412153 | 3.580195 |
| Woodland Hill | 4.295761 | 3.255777 |
| Oakland | 4.382144 | 3.398177 |
| *Average* | *4.618899* | *3.694128* |

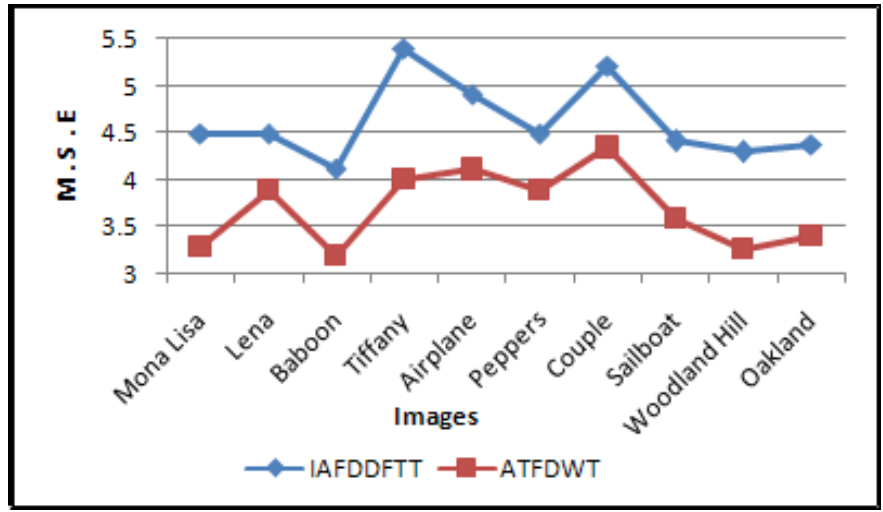

Fig. 9: Graphical representation of MSE of for various benchmark images of ATFDWT in comparison with IAFDDFT

## 4.2 Peak Signal to Noise Ratio (PSNR)

The peak signal-to-noise ratio, often abbreviated PSNR [4], is the ratio between the maximum possible power of a signal and the power of corrupting noise that affects the fidelity of its representation PSNR is usually expressed in terms of the logarithmic decibel scale. On comparison with IAFDDFTT, the proposed ATFDWT provides better performance and on average the value of PSNR is 0.978624 dB more than IAFDDFTT shown in table 3. Graphical representation of the same is shown in fig. 10.

Table 3: PSNR of ten benchmark images in IAFDDFTT and ATFDWT

| IMAGES | PSNR (dB) | |
| --- | --- | --- |
| | IAFDDFTT | ATFDWT |
| Mona Lisa | 41.600073 | 42.954014 |
| Lena | 41.608264 | 42.236749 |
| Baboon | 41.985650 | 43.094628 |
| Tiffany | 40.812837 | 42.110488 |
| Airplane | 41.222393 | 41.992542 |
| Peppers | 41.612029 | 42.234927 |
| Couple | 40.964204 | 41.752647 |
| Sailboat | 41.684298 | 42.591737 |
| Woodland Hill | 41.800402 | 43.004257 |
| Oakland | 41.713938 | 42.818344 |
| *Average* | *41.500409* | *42.479033* |

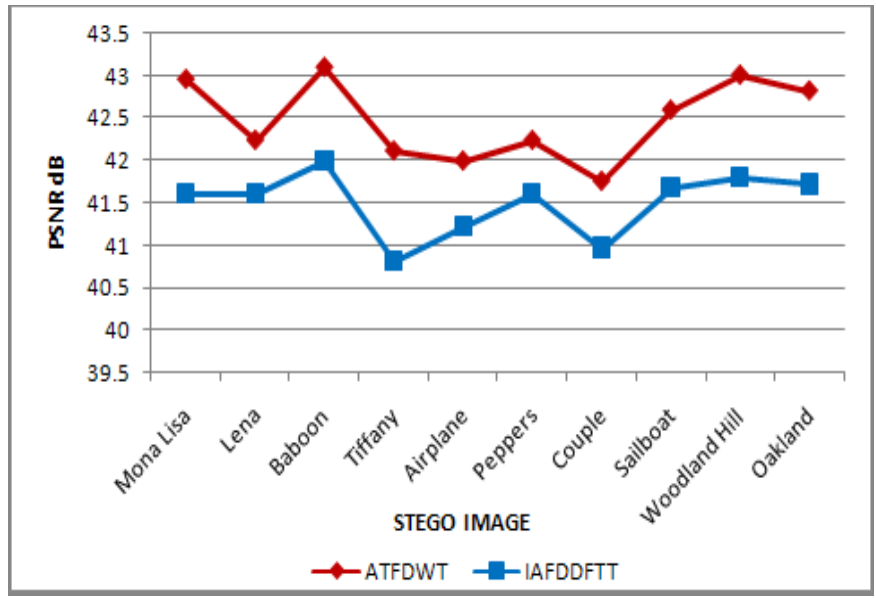

Fig. 10: Graphical representation of PSNR for ten benchmark images in IAFDDFTT and ATFDWT.

### 4.3 Standard Deviation (SD) Analysis

The standard deviation is a measure of the frequency distribution. If an image is uniform throughout, then the standard deviation will be small. On comparison of the difference of SD of the original image and the stego-image, the proposed technique offers better performances. The minimum difference of SD between original and stego-image for existing technique IAFDDFTT

is 0.018029 [Tiffany (44.130630-41.875294)], and min difference for proposed ATFDWT is 0.003735 [Lena (58.964100-58.960365)], maximum difference is 0.144406 for IAFDDFTT and 0.149003 for ATFDWT; Results of SD for each of benchmark image is given in table 4.

Table 4: Comparison of SD of original image and stego-image in IAFDDFTT and ATFDWT

| IMAGES | STANDARD DEVIATION | | |
|---|---|---|---|
| | ORIGINAL | IAFDDFTT | ATFDWT |
| Mona Lisa | 74.826286 | 74.944351 | 74.755318 |
| Lena | 58.964100 | 59.083618 | 58.960365 |
| Baboon | 56.183243 | 56.283375 | 56.093765 |
| Tiffany | 41.875294 | 41.857265 | 41.921524 |
| Airplane | 44.013035 | 44.130630 | 43.992962 |
| Peppers | 66.173019 | 66.230141 | 66.036728 |
| Couple | 32.424759 | 32.526287 | 32.373444 |
| Sailboat | 68.048347 | 68.126648 | 68.040909 |
| Woodland Hill | 43.242332 | 43.386738 | 43.152962 |
| Oakland | 43.483170 | 43.573990 | 43.334167 |
| *Average* | *52.923358* | *53.014304* | *52.866214* |

## 4.4 Image fidelity (IF)

Image fidelity is a parametric computation to quantify the perfectness of human visual perception. On comparison of image fidelity (IF) [4], proposed technique ATFDWT provides better performance than existing IAFDDFTT. The values obtained for IF in ATFDWT and IAFDDFTT is given in table 5 and that of graphical representation of the same is shown in fig.11. Both from the figure and the tabulator data, it is observed that better result is obtained from the proposed ATFDWT for all benchmark images taken for the experiment.

Table 5: Comparison of Image Fidelity between IAFDDFTT and ATFDWT

| IMAGES | IMAGE FIDELITY | |
|---|---|---|
| | IAFDDFTT | ATFDWT |
| Mona Lisa | 0.999749 | 0.999816 |
| Lena | 0.999775 | 0.999805 |
| Baboon | 0.999785 | 0.999834 |
| Tiffany | 0.999874 | 0.999906 |
| Airplane | 0.999860 | 0.999883 |
| Peppers | 0.999730 | 0.999766 |
| Couple | 0.997584 | 0.997985 |
| Sailboat | 0.999778 | 0.999820 |
| Woodland Hill | 0.999820 | 0.999864 |
| Oakland | 0.999796 | 0.999841 |
| *Average* | 0.999575 | *0.999652* |

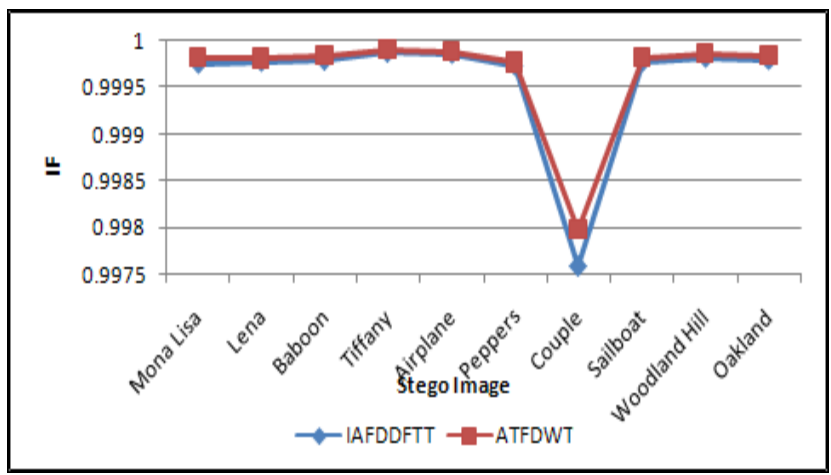

Fig. 11: Graphical representation of IF using IAFDDFTT and ATFDWT.

## 5 Conclusion

In this paper, the issue of hiding data in cover image in frequency domain, through discreet wavelet transformation technique (DWT) has been addressed. On comparison with other standard techniques, it is observed that the proposed ATFDWT obtain better performances.

### Acknowledgement

The authors express deep sense of gratuity towards the department of CSE University of Kalyani and the IIPC Project AICTE, Govt. of India, of the department where the computational resources were used for the work.